\documentclass{article}

\def\xxinput#1{\input#1}

\xxinput{vsolj02.sty}

\usepackage[dvipdfmx]{graphicx}
\usepackage[comma,colon]{natbib}

\usepackage[OT2,T1]{fontenc}

\def\cite{\citealt}
\setcitestyle{aysep={}}

\newcounter{author}
\def\altaffilmark#1{$^{#1}$}
\def\altaffiltext#1{$^{#1}$\,}

\def\authorcount#1#2{{\refstepcounter{author}\label{#1}
                     \altaffiltext{\ref{#1}}{#2}}}

\begin{document}

\begin{center}

\title{Low state in the post-nova V1315 Aql}

\author{
        Taichi~Kato\altaffilmark{\ref{affil:Kyoto}}
}

\authorcount{affil:Kyoto}{
     Department of Astronomy, Kyoto University, Sakyo-ku,
     Kyoto 606-8502, Japan \\
     \textit{tkato@kusastro.kyoto-u.ac.jp}
}

\end{center}

\begin{abstract}
\xxinput{abst.inc}
\end{abstract}

   V1315 Aql was initially discovered as a variable star
(=SVS 8130) by \citet{met61v1315aql}.
During a search for ultraviolet-excess objects,
\citet{dow86v1315aql} detected this object (KPD 1911$+$1212)
and identified it to be an eclipsing cataclysmic variable.
Its spectrum was that of a high-excitation old nova0
(this term was used slightly differently from the modern
meaning), and unusually singly peaked emission lines despite
its high inclination and the unusual behavior of the emission
lines against the orbital phase already attracted attention.
\citet{dow86v1315aql} classified V1315 Aql to be a member
of high-excitation (all eclipsing) old novae, including SW Sex,
LX~Ser, DQ Her, RW Tri and V363 Aur as a group.
These unusual emission lines have been studied by many authors
\citep{szk87ippegdwumav1315aqlIUE,szk90v1315aqlswsexdwuma,
dhi91v1315aql,smi93v1315aql,dhi95v1315aql,hel96v1315aql}.
This object played an important role in establishing
the concept of SW Sex stars \citep{tho91pxand,hel96v1315aql}.

   This object received attention again by the discovery
of a nova shell \citep{sah15novashell,sah18v1315aql}.
Using the expansion velocity, \citet{sah18v1315aql}
suggested that the nova eruption occurred 500--1200~yr ago.
\citet{sah18v1315aql} could not find a corresponding
``guest star'' in old Chinese and Asian records compiled
by \citet{ste76historicalnovae}.
\citet{sch19v1315aqlj2102atel12474} could not find evidence
for a nova eruption in archival plates starting from 1889.

   While inspecting of Asteroid Terrestrial-impact
Last Alert System (ATLAS: \cite{ATLAS}) forced photometry
\citep{shi21ALTASforced} and Zwicky Transient Facility
(ZTF: \cite{ZTF})\footnote{
   The ZTF data can be obtained from IRSA
$<$https://irsa.ipac.caltech.edu/Missions/ztf.html$>$
using the interface
$<$https://irsa.ipac.caltech.edu/docs/program\_interface/ztf\_api.html$>$
or using a wrapper of the above IRSA API
$<$https://github.com/MickaelRigault/ztfquery$>$.
} data, I noticed that this object entered a low state
starting from early 2023 (vsnet-alert 27760)\footnote{
   $<$http://ooruri.kusastro.kyoto-u.ac.jp/mailarchive/vsnet-alert/27760$>$.
}.  As far as I know, this is the first such an event
since the discovery of this object.

\begin{figure*}
\begin{center}
\includegraphics[width=16cm]{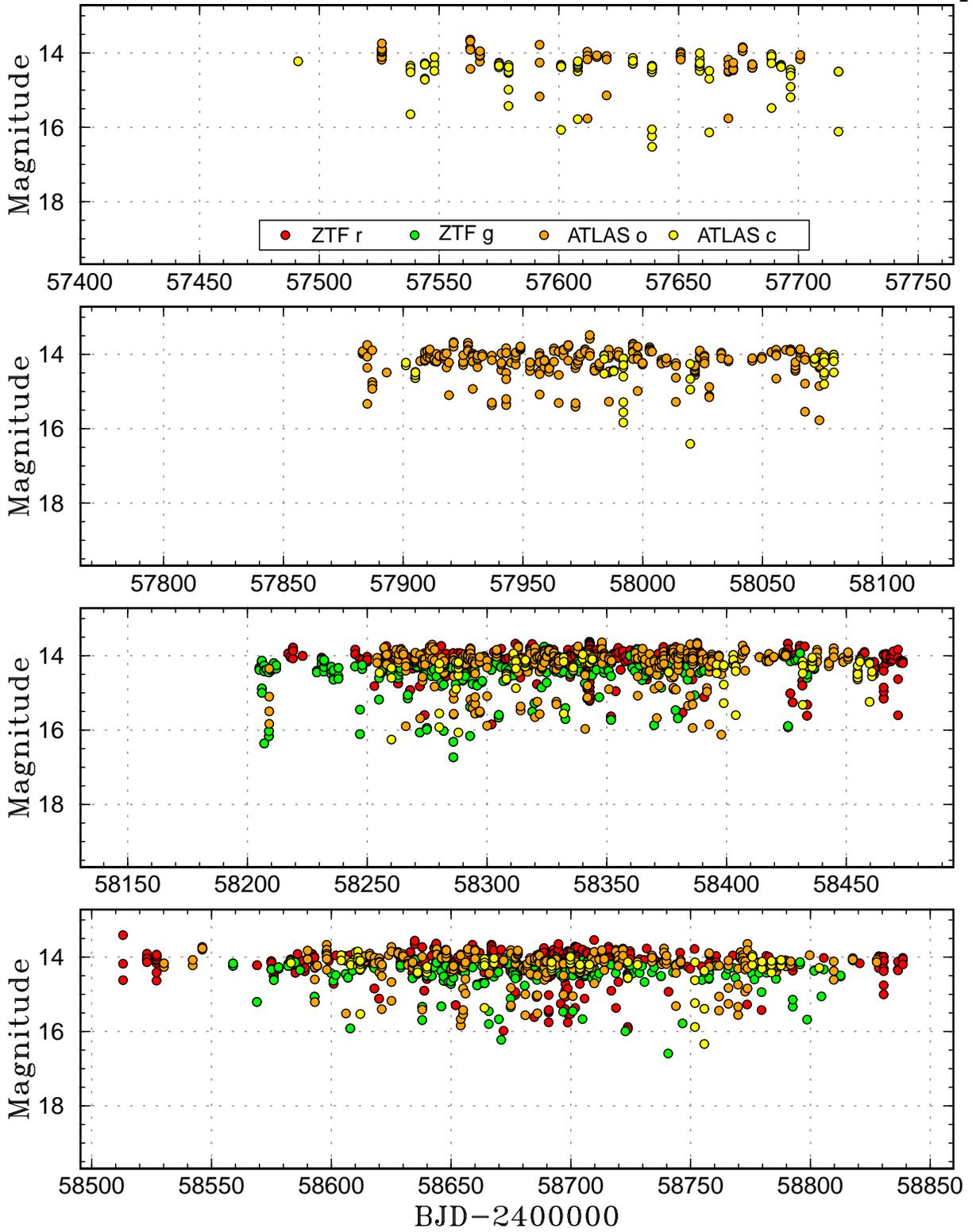}
\caption{
   Light curve of V1315 Aql in 2016--2019.
The object was in high state.  Eclipses with depths of
1--2~mag were also present.
}
\label{fig:lc1}
\end{center}
\end{figure*}

\begin{figure*}
\begin{center}
\includegraphics[width=16cm]{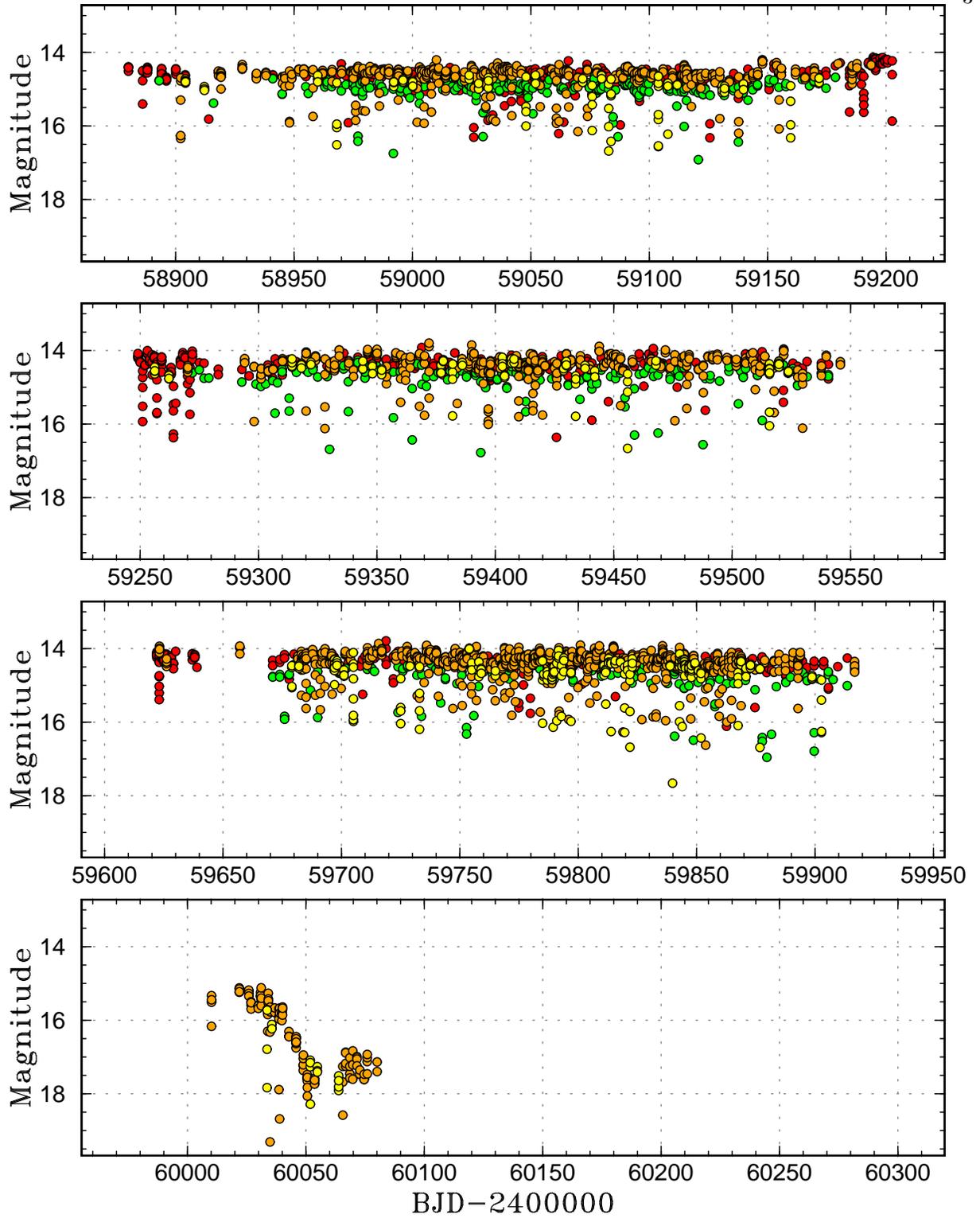}
\caption{
   Light curve of V1315 Aql in 2020--2023.
The symbols are the same as in figure \ref{fig:lc1}.
The object is in now low state in 2023 (fourth panel).
The bottoms of eclipses in the low state were probably
too faint to detect by ATLAS.
}
\label{fig:lc2}
\end{center}
\end{figure*}

   The typical high-state light curve (2016--2019) is shown
in figure \ref{fig:lc1}.  The object gradually faded
and entered a low state starting from 2023
(figure \ref{fig:lc2}).
\citet{hel00swsexreview} suggested that SW Sex
behavior is caused by episodes of very high mass transfer,
which are balanced by VY Scl low states
(i.e. to adjust to the long-term mass transfer by angular
momentum loss from the binary).  In the case of
V1315 Aql, the origin of very high mass transfer in its
high state appears to be the result of a nova explosion.
The present low state would provide an opportunity to
study the secondary (such as the chemical composition)
in detail, which has been hampered by the luminous
accretion disk in the past.

\begin{figure*}
\begin{center}
\includegraphics[width=16cm]{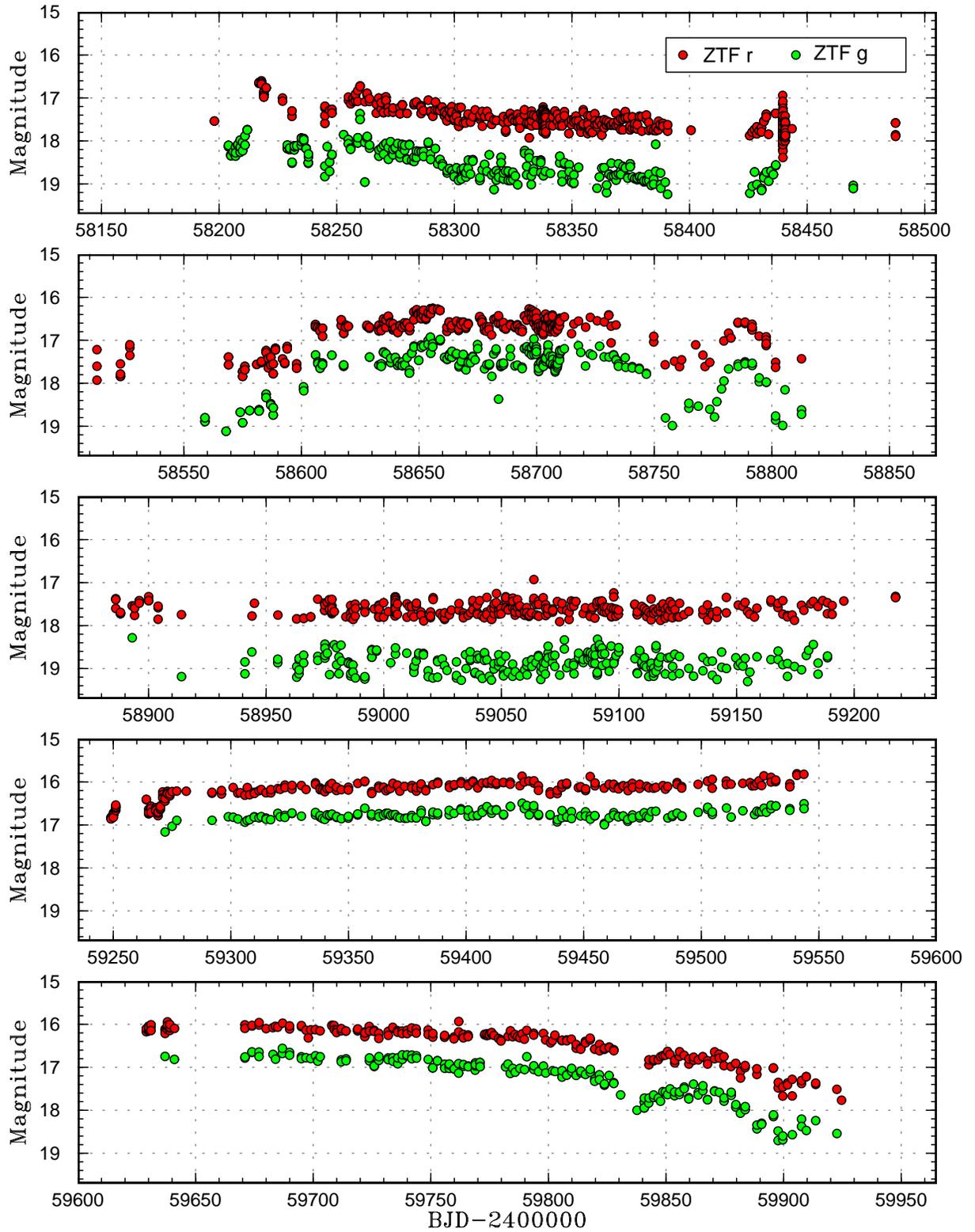}
\caption{
   Light curve of LV Vul in 2018--2022.
Low state (third panel) and high state (fourth panel);
transition between these state in other panels.
}
\label{fig:lvvullc1}
\end{center}
\end{figure*}

   Several classical novae (not necessarily SW Sex stars)
have been known to show low states recently.
They are summarized in table \ref{tab:novalow}.
For CP Lac see vsnet-alert 23621\footnote{
   D. Denisenko
   $<$http://ooruri.kusastro.kyoto-u.ac.jp/mailarchive/vsnet-alert/23621$>$.
}, which was originally detected by Gaia.\footnote{
   $<$http://gsaweb.ast.cam.ac.uk/alerts/alert/Gaia19elx/$>$.
}
The 2020--2022 low state of DK Lac was detected in the ZTF data.
The light curve of LV Vul from the ZTF data is shown
in figure \ref{fig:lvvullc1}.
In addition to them, there are also more candidates:
\begin{itemize}
\item
IV Cep (1971): highly variable in the ZTF data.  More than 1~mag
fading in 2018 December--2019 April, 2020 June--August,
2021 September--? (ending not recorded due to the seasonal gap)
and 2022 August--September.
\item
V400 Per (1974): likely low state in early 2022; only three
positive ZTF observations.
\end{itemize}

\begin{table*}
\caption{Classical novae with low states.}
\label{tab:novalow}
\begin{center}
\begin{tabular}{cccc}
\hline
Object & Eruption & Low states & References \\
\hline
CP Lac & 1936 & 2019--2020 & vsnet-alert 23621 \\
DK Lac & 1950 & 2000--2001, 2020--2022 & \citet{hen01dklaciauc7777} \\
HR Lyr & 1919 & 2010 (shallow) & \citet{hon14hrlyr} \\
LV Vul & 1968 & 2018, 2020, 2022 & this work \\
V1315 Aql & before 1889 & 2022 & this work \\
\hline
\end{tabular}
\end{center}
\end{table*}

\section*{Acknowledgements}

This work was supported by JSPS KAKENHI Grant Number 21K03616.
I am grateful to Naoto Kojiguchi for helping downloading
ZTF data.

This work has made use of data from the Asteroid Terrestrial-impact
Last Alert System (ATLAS) project.
The ATLAS project is primarily funded to search for
near earth asteroids through NASA grants NN12AR55G, 80NSSC18K0284,
and 80NSSC18K1575; byproducts of the NEO search include images and
catalogs from the survey area. This work was partially funded by
Kepler/K2 grant J1944/80NSSC19K0112 and HST GO-15889, and STFC
grants ST/T000198/1 and ST/S006109/1. The ATLAS science products
have been made possible through the contributions of the University
of Hawaii Institute for Astronomy, the Queen's University Belfast, 
the Space Telescope Science Institute, the South African Astronomical
Observatory, and The Millennium Institute of Astrophysics (MAS), Chile.

Based on observations obtained with the Samuel Oschin 48-inch
Telescope at the Palomar Observatory as part of
the Zwicky Transient Facility project. ZTF is supported by
the National Science Foundation under Grant No. AST-1440341
and a collaboration including Caltech, IPAC, 
the Weizmann Institute for Science, the Oskar Klein Center
at Stockholm University, the University of Maryland,
the University of Washington, Deutsches Elektronen-Synchrotron
and Humboldt University, Los Alamos National Laboratories, 
the TANGO Consortium of Taiwan, the University of 
Wisconsin at Milwaukee, and Lawrence Berkeley National Laboratories.
Operations are conducted by COO, IPAC, and UW.

The ztfquery code was funded by the European Research Council
(ERC) under the European Union's Horizon 2020 research and 
innovation programme (grant agreement n$^{\circ}$759194
-- USNAC, PI: Rigault).

\section*{List of objects in this paper}
\xxinput{objlist.inc}

\section*{References}

We provide two forms of the references section (for ADS
and as published) so that the references can be easily
incorporated into ADS.

\newcommand{\noop}[1]{}\newcommand{\hyphalt}{-}

\renewcommand\refname{\textbf{References (for ADS)}}

\xxinput{v1315aqlaph.bbl}

\renewcommand\refname{\textbf{References (as published)}}

\xxinput{v1315aql.bbl.vsolj}

\end{document}